# High-pressure synthesis of the indirectly electron-doped 122 iron superconductor $Sr_{1-x}La_xFe_2As_2$ with a maximum $T_c$ = 22 K

Yoshinori Muraba[1], Satoru Matsuishi[1], Sung-Wng Kim[2], Toshiyuki Atou[1], Osamu Fukunaga[2] and Hideo Hosono[1,2,]*

[1] Materials and Structures Laboratory, Tokyo Institute of Technology, 4259 Nagatsuta, Midori, Yokohama 226-8503, Japan

[2] Frontier Research Center, Tokyo Institute of Technology, 4259 Nagatsuta, Midori, Yokohama 226-8503, Japan

Compounds of $Sr_{1-x}La_xFe_2As_2$ were synthesized by solid state reaction at 1273 K, under pressures of 2 or 3 GPa. The $Sr_{1-x}La_xFe_2As_2$ phase was dominant up to $x$ = 0.5, and superconductivity was observed at $x \geq 0.2$. A maximum critical temperature $T_c$ of 22 K, and a maximum shielding volume fraction of ~70 % were obtained at $x$ = 0.4. This is the first experimental demonstration of electron-doped superconductivity in 122-type iron pnictides, where electrons were doped through aliovalent substitution at the site of the alkaline earth metal. The optimal $T_c$ was slightly higher than that for the Co-doped (directly electron-doped) samples (19 K), and much lower than that for the hole-doped case (37 K). The bulk superconductivity range was narrower than that for the Co-substituted case, and both ranges were much narrower than that for the hole-doped case. These observations revealed that differences in the electron-doping mode (direct or indirect) did not have a prominent effect on the optimal $T_c$ or superconductivity range, compared with differences in carrier polarity.

*Corresponding author: Hideo Hosono
Materials and Structures Laboratory, Tokyo Institute of Technology
4259 Nagatsuta, Midori, Yokohama 226-8503, Japan
TEL +81-45-924-5359
FAX +81-45-924-5339
E-mail: hosono@msl.titech.ac.jp

The discovery of $LaFeAsO_{1-x}F_x$ with a critical temperature ($T_c$) of 26 K in early 2008 triggered the extensive exploration for new iron-based superconductors.[1] Various compounds with an iron square lattice have since been reported, and the $T_c$ was pushed up to 56 K. This exceeded the maximum $T_c$ of other materials, with the exception of that for high $T_c$ cuprates.[2-4] Such superconductors have a 2-dimensional cylindrical Fermi surface with 2 hole pockets at the Γ-point and 2 electron pockets at the M point. The origin of the superconductivity is usually discussed in terms of nesting on the Fermi surface.[5,6] Angle-resolved photoemission spectroscopy (ARPES) experiments[7] and theoretical calculations[5,6] suggest that superconductivity is expected to occur by electron or hole doping, providing the hard band model is valid. The experimental examination of the both electron-doped and hole-doped case of superconductivity is still far from complete. For example, in the 1111 phase it is established that electron-doping effectively works to induce superconductivity as exemplified by $Ln$$FeAsO_{1-x}F_x$ (where *Ln*: lanthanide). However, the hole doping effect still remains unclear.[8] Information on both electron-doped and hole-doped cases in superconductivity is important for examining the modification of the Fermi surface by carrier doping.

The 122 phase[9] is more favorable than the 1111 phase for examining this issue, because the phase forming reaction temperature is sufficiently low that sub-reactions within container vessels and/or selective vaporization of a component may be suppressed. Direct electron-doping to FeAs-layers by substituting the iron site with appropriate transition metals (e.g. $Co^{2+}$, $Ni^{2+}$)[10,11] can be achieved and superconductivity was induced by the doping mode. However, to date there are no reports of superconductivity of 122 system induced by indirect electron-doping via aliovalent cation substitution to alkaline earth metal site. While G. Wu et al.[12] examined the $La^{3+}$ substitution at the alkaline earth metal site, solid state reactions of the ingredient mixture for $Ba_{1-x}La_xFe_2As_2$ did not yield the La-substituted 122 phase by using the conventional glass-tube technique under ambient pressure. Leithe-Jasper et al.[13] reported density functional theory (DFT) calculations showing a distinct difference in phase instability and effect of magnetic moment, originating from the electronic structure between Co-doped and La-doped $SrFe_2As_2$. The La-substituted phase was electronically unstable, but no phase instability was induced by Co-doping. This prediction is consistent with the formation of $Sr_{1-x}La_xFe_2As_2$ phases having not been reported to date.

The ionic radius of $La^{3+}$ (116 pm) is smaller than that of $Sr^{2+}$ (126 pm), and so it is likely that a high pressure synthesis would be effective for obtaining La-substituted $SrFe_2As_2$. In the current study, we report that $Sr_{1-x}La_xFe_2As_2$ ($x \leq 0.6$) compounds with negative Seebeck coefficients can be synthesized at high pressure (2-3 GPa), and superconductivity with a maximum $T_c$ of 22 K was observed in the range of $0.2 \leq x \leq 0.5$.

Samples of nominal chemical composition $Sr_{1-x}La_xFe_2As_2$ were synthesized by solid state reaction under high pressure. The starting materials were SrAs, LaAs and $Fe_2As$, each of them having been prepared from their respective metals. The mixture of these materials was sealed in a metal Ta tube with a Ta cap, by spot-welding in a glove box filled with a dry Ar gas. Tubes were then heated at 1273 K for 2 h under pressures of 2-4 GPa. A belt-type anvil cell was employed for the high pressure synthesis. Crystalline phases in the resulting samples were identified by powder X-ray diffraction (XRD) using a Bruker diffractometer model D8 ADVANCE (Cu rotating anode). The Rietveld analysis of XRD patterns was performed using RIETAN-FP code[14]. The 4-probe resistivity and magnetic susceptibility were measured in the temperature range of 300-2 K, using a PPMS (Quantum Design) with a VSM attachment. Thermopower was measured at 300 K to check the polarity of the major carrier in the resulting samples.

Because solubility of La in $AeFe_2As_2$ ($Ae$ = Sr, Ba) is small, solid state reactions of the ingredient mixture for $Ae_{1-x}LaxFe_2As_2$ did not yield the La-substituted 122 phase by using the conventional glass-tube technique under ambient pressure. LaAs and $SrFe_2As_2$ phases were identified by XRD, indicating segregation of La from the 122 phase. When a pressure of 2 GPa was applied during the reaction, the formation of LaAs was suppressed and an XRD peak shift was observed. When the applied pressure was increased to 3 GPa, the volume fraction of impurities increased. The target became a minority phase at 4 GPa, and so reaction conditions were fixed at 2 GPa and 1273 K for 2 h.

Figure 1 (a) shows observed and calculated XRD patterns of samples synthesized at 2 GPa. Segregation of La was suppressed up to $x = 0.5$, but the LaAs phase was apparent for $x = 0.6$. A trace amount of LaFeAsO (< 3 vol %) was identified in all samples. The same amount of LaFeAsO phase was observed as for samples prepared by the glass-tube technique, thus the formation of LaFeAsO was attributed to oxygen contamination in ingredients during preparation and synthesis. Figure 1 (b) shows the variation of lattice constants ($a$ and $b$), unit cell volume $V$, and the separation between Sr and As ions $r_{\text{Sr-As}}$ within the $Sr_{1-x}La_xFe_2As_2$ phase as functions of nominal $x$. The unit cell volume continuously decreased from $x = 0$ to 0.5, indicating that the solubility limit of La in $SrFe_2As_2$ was $x < 0.6$. Providing there was separation between Sr and As ions in the 122 phase, $r_{\text{Sr-As}}$, decreased with the mean size of the cations at the Sr-site, and the value of $r_{\text{Sr-As}}$ should decrease linearly with $x$. The difference in $r_{\text{Sr-As}}$ between $x = 0$ and $x = 1$ (extrapolated) should be the difference in the ionic radius between $Sr^{2+}$ and $La^{3+}$ (~10 pm). The observed trend and value (~10 pm) in Figure 1 (b) agrees with the above assumption, confirming that $La^{3+}$ substitutes the site of $Sr^{2+}$. With an increase of nominal $x$, the broadening of diffraction peaks increased, especially for the 00$l$ diffractions. The length of the $c$-axis was more sensitive to $x$ variations than that of the $a$-axis (See Figure 1 (b)). A possible reason for the broadening was the modulation of lattice spacing, due to inhomogeneity in local

La-concentration. The modulation of lattice spacing yielded the crystallographic strain. The degree of modulation as crystallographic strain was calculated from the line-broadening obtained by the Rietveld analysis of the XRD patterns. Figure 1 (c) shows isotropic and anisotropic crystallographic strains as a function of *x*. $S_G$ and $S_L^{iso}$ are the isotropic strains calculated from Lorentzian and Gaussian-type components of pseudo-Vogt functions, which describe line broadening for the Rietveld analysis. $S_L^{aniso}$ is the anisotropic strain along the [001] axis, calculated from the Lorentzian component.[15] In this analysis, the value of parameter corresponding to crystallite size was fixed to that optimized for the sample of $x = 0$. Both isotropic and anisotropic lattice strains start to increase from $x = 0.1$, and then increase markedly when $x > 0.4$. Thus, it is presumed that the large strain induced in the crystallites by the La-substitution restricted the solubility limit of $La^{3+}$ at the $Sr^{2+}$ site.

Figure 2 shows the temperature dependence of resistivity for samples with different *x*. An anomaly arising from the crystallographic transition[16] accompanied the magnetic transition at around 200 K, and was suppressed with *x*. A sharp resistivity drop instead appeared from $x = 0.2$, and zero resistivity was observed at $x \geq 0.3$.

Figure 3 shows the temperature dependence of magnetic susceptibility $\chi$ for each sample. Pellets were obtained by sintering synthesized powders, and were used for the measurements. Samples all underwent demagnetization at 300 K before magnetic measurements were carried out. A distinct negative shift of $\chi$ was observed for $x = 0.2$-$0.5$ in the temperature range below ~22 K. The top of Figure 3 (b) shows *M-H* curves for the samples. The shielding volume fraction calculated from the slope of the *M-H* curve was ~10 % at $x = 0.3$, and reached a maximum of ~70 % at $x = 0.4$. It then dropped to ~5 % at $x = 0.5$. The precipitation of impurity phases was apparent from XRD patterns for the samples prepared at 3GPa, indicating that the La-concentration in the Sr-122 crystals was smaller than the nominal *x*. Since the *c*-axis length for samples prepared at 2 GPa was almost proportional to the nominal *x*, it could be used as a measure for the content of La substituting at the Sr sites. Figure 3 (c) shows the shielding volume fraction evaluated from *dM/dH* vs. lattice constant *c* for samples prepared at 2 and 3 GPa. The samples appeared to also display a maximum volume fraction of superconducting phase at the same *c*-axis length.

Figure 4 shows the Seebeck coefficients of $Sr_{1-x}M_xFe_2As_2$($M$ = K and La) at 300 K. It was evident that the sign of Seebeck coefficients for the La-doped samples was negative, and opposite to that observed for the K-doped samples.[17] This result substantiated that electrons were effectively doped to the FeAs layer by the La-substitution as intended, and that superconductivity was induced by electron-doping.

Analysis of XRD patterns indicated that La-substitution induced anisotropic strain along the *c*-direction. Saha et al. reported that the superconductivity of non-doped $SrFe_2As_2$ was promoted by internal strain, and caused by stress during crystal growth which disappeared during thermal annealing.[18] In the current study, we examined the annealing at 573 K of $Sr_{0.6}La_{0.4}Fe_2As_2$ prepared at 2 GPa, but found no significant changes in superconducting properties ($T_c$ and shielding volume) or XRD profile. This indicated that the superconductivity of our samples was not due to mechanical stress like that reported by Saha et al.

The reason for $T_c$ on the ρ-$T$ profile not changing with La concentration was then investigated. The profile analysis of XRD patterns revealed that there is no indication of phase separation but distinct persistent crystalline strain are present in the (Sr,La)$Fe_2As_2$. This strain is attributed to inhomogeneous replacement of Sr ions with La ions in the $SrFe_2As_2$ phase. In each sample with the observable $T_c$, there was a portion in which the La-concentration was the same as $x$ = 0.4 with a maximum $T_c \sim 22$ K and these portions are percolated throughout the bulk samples. Since the onset $T_c$ is controlled by the highest-$T_c$ region ($La_{0.4}Sr_{0.6}Fe_2As_2$), we may understand the obtained results that the shielding volume fraction changed with $x$, but $T_c$ remained almost unchanged. Here, note that substitution between Sr and La is easy in cuprates even in conventional solid state reaction processes but impossible in Sr-122 iron arsenide. This makes a sharp contrast between these two representative superconductive systems

Figure 5 summarizes the electronic phase diagram for directly/indirectly, hole/electron-doped $SrFe_2As_2$. Onset-$T_c$ and anomaly temperature in ρ-$T$ curves indicating structural/magnetic transition are plotted in this figure. Here, doped carrier numbers per Fe are plotted in place of compositions in the figure. It is of interest to compare the present results (indirect electron-doping) with those for $SrFe_{2-y}Co_yAs_2$ (direct doping).[13,16] Superconductivity in the latter system occurred at $0.2 \leq y \leq 0.5$. $T_c$ had a maximum of 19 K at $y \sim 0.2$, at which point anti-ferromagnetic ordering decreased with $x$ and disappeared at $y$ = 0.5. In the present system, bulk superconductivity (shielding volume > ~10 %) was observed at $x$ = 0.3 and 0.4 for samples prepared at 2 GPa, and $x$ = 0.4 and 0.5 for those prepared at 3 GPa. The number of electrons injected into the FeAs layer was $x/2 = y$ per Fe site. Superconductivity was induced for the La-substituted system if electron numbers of 0.15-0.25 per Fe for samples prepared at 2 GPa, and 0.2-0.25 per Fe site for those prepared at 3 GPa, were doped to the parent phase. However, the $T_c$ remains almost constant (22 K) irrespective of $x$ and the shielding volume fraction is very small except at $x$ = 0.4. Thus, it is considered that superconducting range in this system is restricted to $x$ = 0.4 ($\Delta n_{electron}$/Fe = 0.2). This range is narrower than that for the Co-substituted system, and both ranges are much narrower than that for the hole-doped case. In this case the range continues to the end member $KFe_2As_2$. The maximum $T_c$ for the indirect (La)-doped system was slightly higher than that (19 K) for the direct(Co)-doped system,[13] but was lower than that (37 K) for the optimal hole-doped case.[17]

Recently, asymmetry in the electronic phase diagram upon electron and hole doping into Fe pnictide superconductors, was theoretically studied by Ikeda et al.[19] They constructed a five-band model using a combination of ab initio band calculations and fluctuation exchange (FLEX) approximations, and examined the doping dependence of superconductivity. They found that superconductivity was stable over a wider range for hole doping than for electron doping. Electron doping filled the hole pockets, which in turn reduced the nesting between the hole pockets and electron pockets. The disappearance of the electron pockets did not occur by hole doping. The present experimental results on the carrier doped 122 system are consistent with this theoretical prediction.

In summary, indirectly electron-doped $Sr_{1-x}La_xFe_2As_2$ was synthesized by solid state reaction under pressures of 2-3 GPa. The optimal $T_c$ was slightly higher than that for the indirectly electron (Co)-doped case, but much lower than that for the hole-doped case. No significant difference in the superconductivity range was observed between La- and Co-substitution. Both ranges were much narrower than that for the hole-doped case. It was concluded that the difference in electron-doping mode, either direct or indirect, was much smaller than that of polarity of the doped carrier.

**Acknowledgments**

This study was supported by the Funding Program for World-Leading Innovative R&D on Science and Technology, JSPS, Japan. The authors thank Drs. T. Kamiya and H. Ikeda for their discussion.

**Figure Captions**

**Fig.1**: X-ray diffraction patterns and crystallographic data. (a) Powder X-ray diffraction (XRD) patterns of $Sr_{1-x}La_xFe_2As_2$ with $x$ = 0.0-0.6. A small amount of LaFeAsO phase denoted by asterisk (volume fraction ~ 3 %) almost independent of $x$, was seen for $x$ = 0.1-0.5 samples. The remarkable line broadening of 00*l* reflections was observed upon La-substitution, indicating the presence of anisotropic strain along the *c*-axis. (b) Lattice constants *a* and *c*, unit cell volume *V*, and Sr-As distance $r_{Sr\text{-}As}$ as a function of *x*: Data on the sample with $x$ = 0 synthesized at an ambient pressure are indicated by the filled circles. (c) Crystallographic strains $S_G$, $S_L^{iso}$ and $S_L^{aniso}$ as a function of *x*: $S_G$ and $S_L^{iso}$ are isotropic strains calculated from Lorentzian and Gaussian-type components of pseudo-Vogt function describing line broadening for Rietveld analysis. $S_L^{aniso}$ is the anisotropic strain along the [001] direction, also calculated from a Lorentzian-type component.

**Fig.2.** DC resistivity of $Sr_{1-x}La_xFe_2As_2$ as a function of temperature. The inset is the enlarged region around $T_c$.

**Fig3.** Magnetic properties of $Sr_{1-x}La_xFe_2As_2$. (a) Temperature dependence of magnetic susceptibility $\chi$ of the samples subjected to zero field cooling (bold trace) or field cooling. (dotted trace). (b) Magnetic moment vs magnetic field at 2 K. Bottom is the shielding volume fraction evaluated from dM/dH vs lattice constants *c* of the samples prepared under pressures of 2 and 3 GPa. The shielding volume fraction trend was almost the same for both series of samples.

**Fig.4.** Seebeck coefficients at 300 K in aliovalent cation-substituted $SrFe_2As_2$. Data on the K-substituted samples were taken from ref.16.

**Fig.5.** Electronic phase diagrams for hole/electron-doped $SrFe_2As_2$. This diagram was drawn based on the data reported in refs. 12, 16 and present work. $\Delta n_{electron}$/Fe is the injected number of electrons per Fe site and $\Delta n_{hole}$/Fe is the number of holes. Observed data (□) on $T_c$ in $Sr_{1-x}La_xFe_2As_2$ is almost constant ~22 K for $\Delta n_{electron}$/Fe = 0.1-0.25. Thus, we attribute intrinsic superconducting region to a narrow range around 0.2 on the basis of the presence of persistent crystallographic strain, shielding volume fractions and onset $T_c$-$x$ relation.

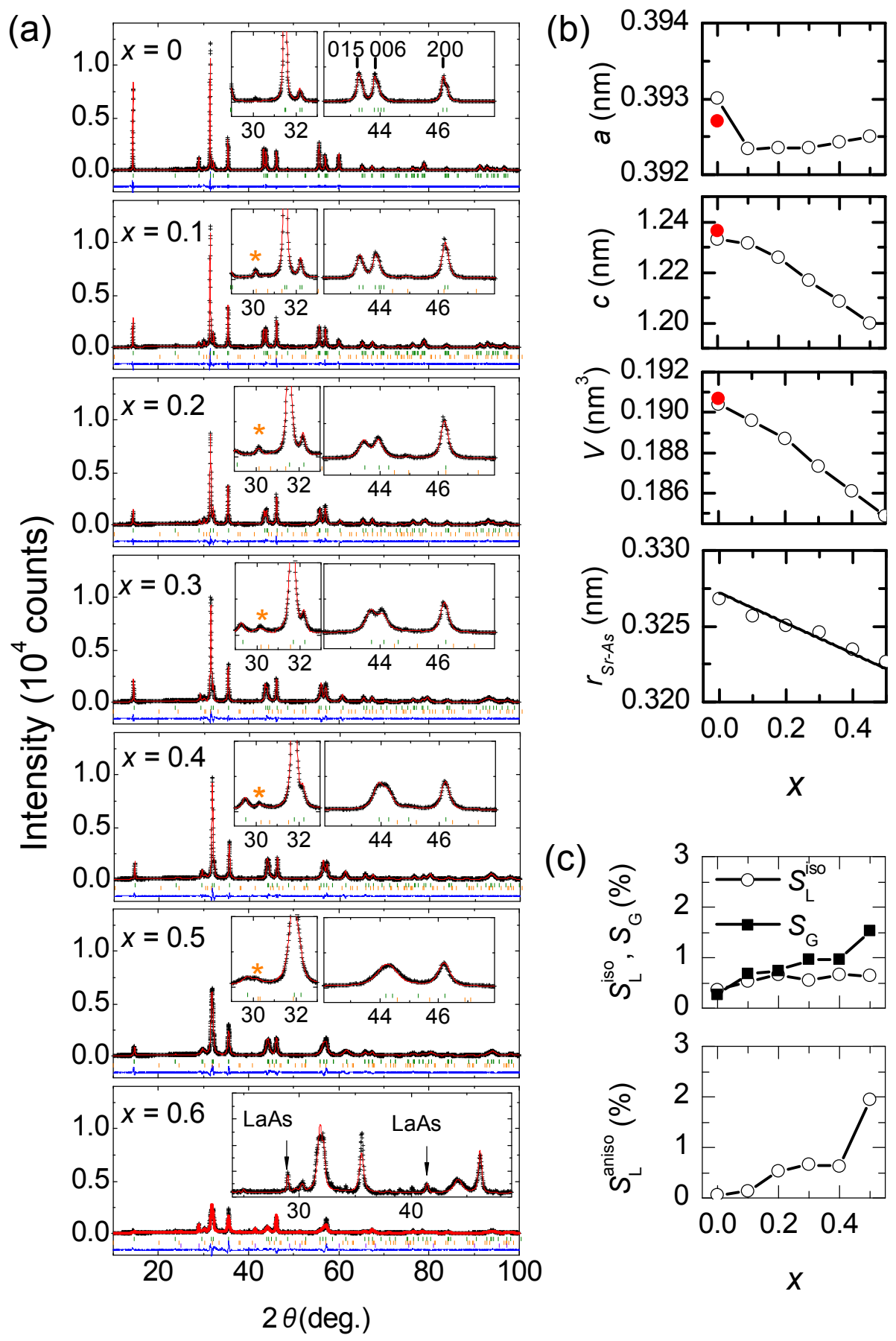

Fig. 1

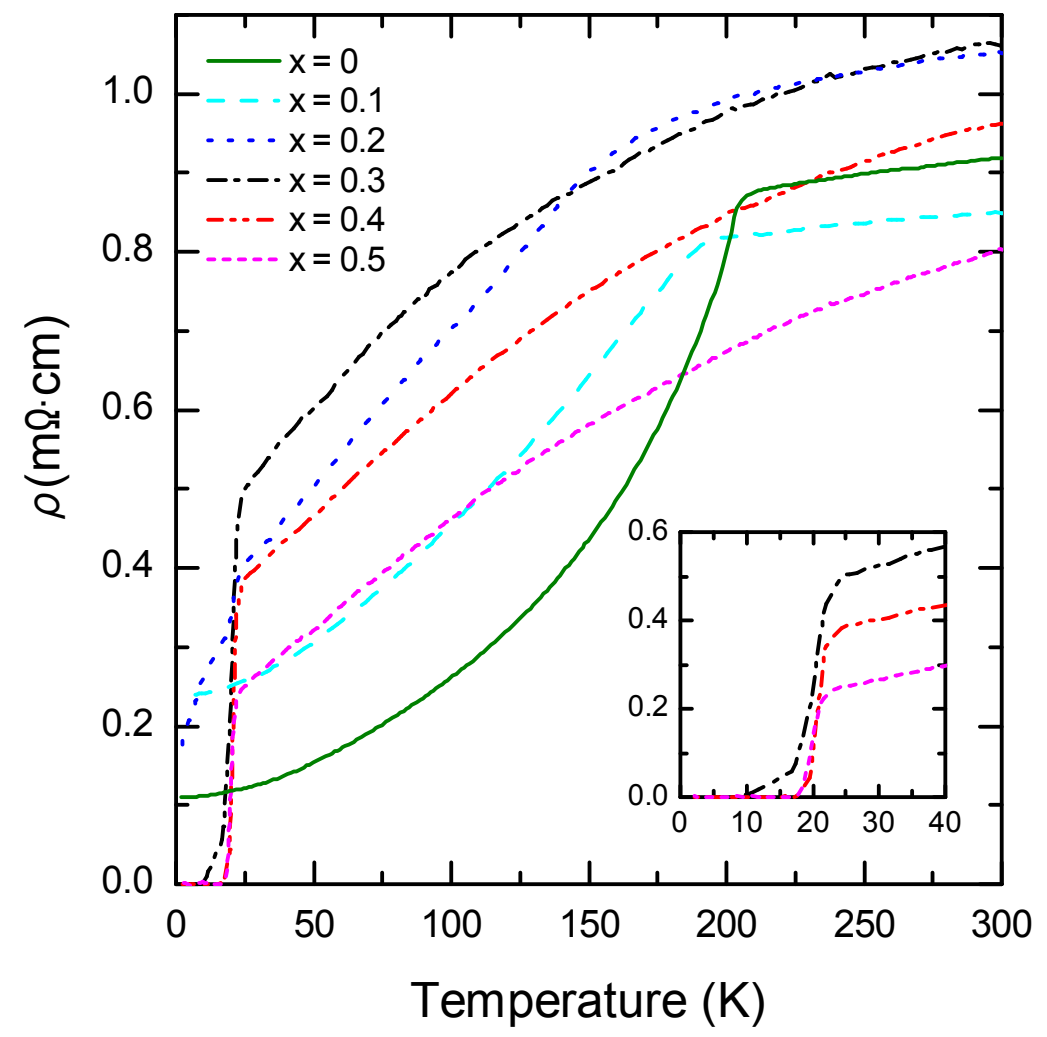

Fig. 2

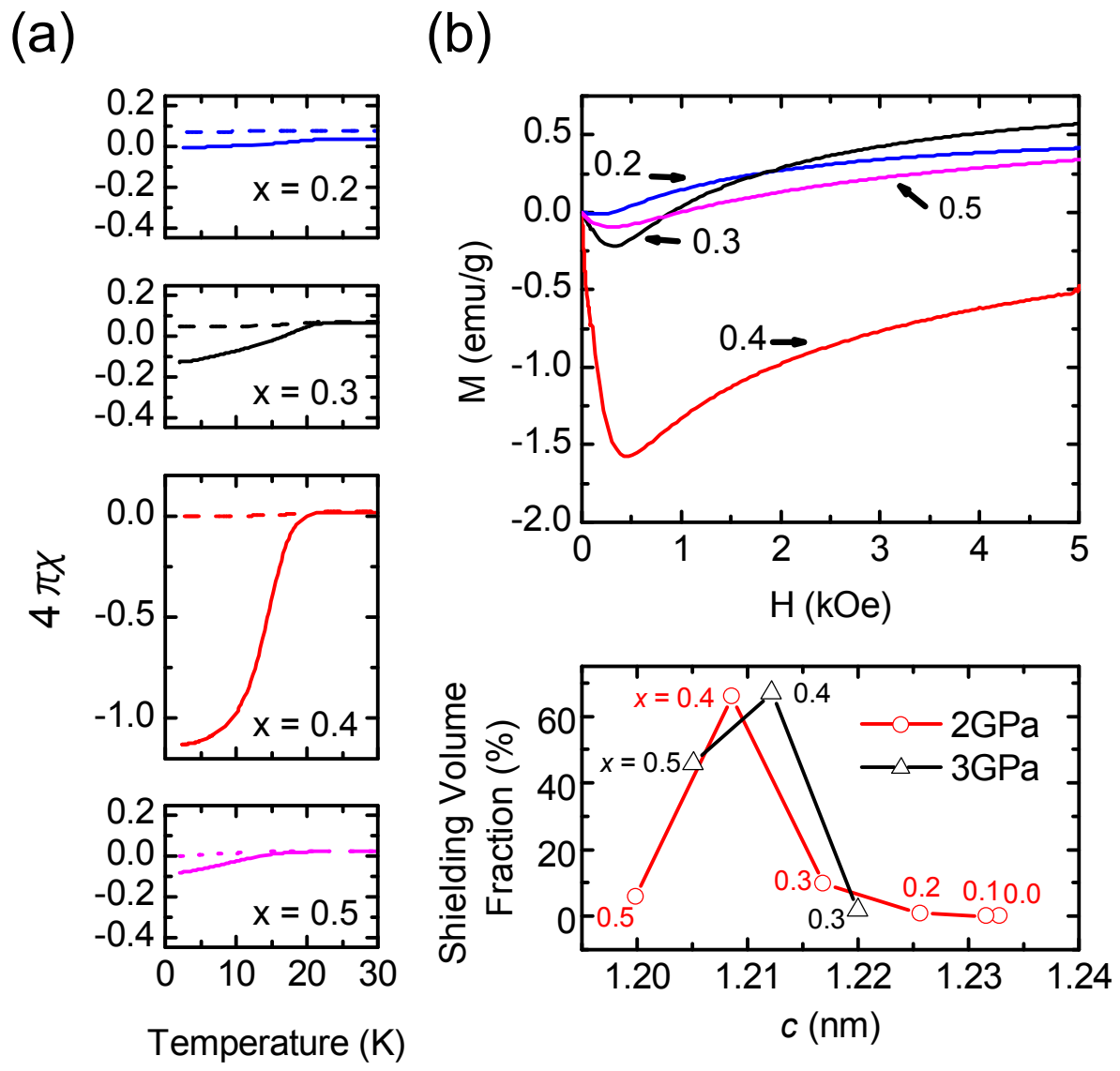

Fig. 3

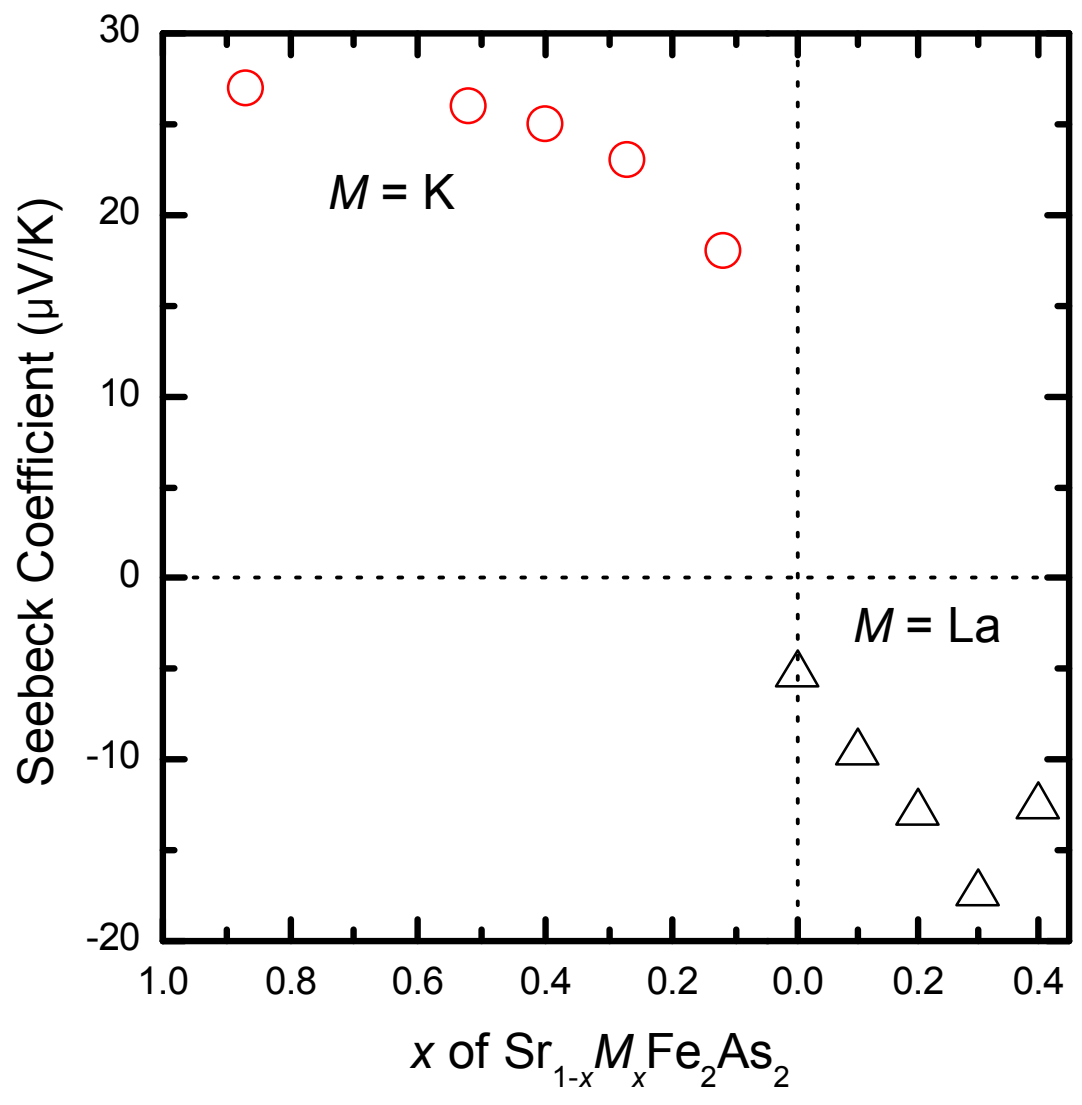

Fig. 4

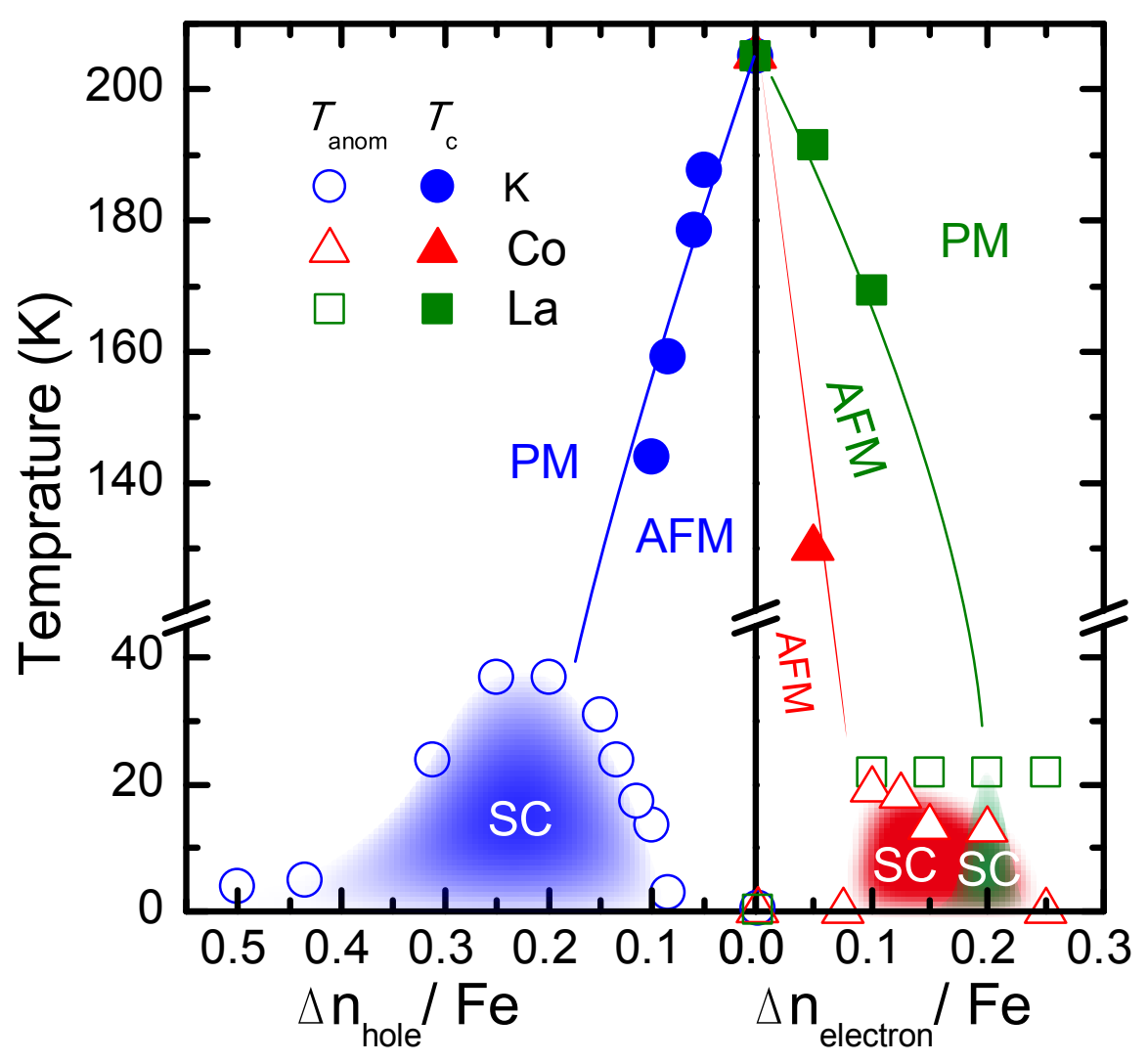

Fig. 5